\documentclass[runningheads]{llncs}

\usepackage[T1]{fontenc}
\usepackage{amsmath,amssymb}
\usepackage{xspace}
\usepackage{tabularx}
\usepackage{booktabs}
\usepackage{multirow}
\usepackage{makecell}

\usepackage{placeins}
\usepackage{siunitx}
\usepackage{url}
\usepackage[shortcuts]{extdash}
\usepackage{graphicx}
\usepackage{nicefrac}
\usepackage{tikz}
\usepackage[super]{nth}
\usepackage{todonotes}
\usepackage{flushend}

\usepackage{tablefootnote}

\usepackage{graphicx}

\newcommand{\fb}[0]{FluencyBank\xspace}
\newcommand{\sep}[0]{SEP\=/28k\xspace}
\newcommand{\sepE}[0]{SEP\=/28k\=/E\xspace}
\newcommand{\ksof}[0]{KSoF\xspace}
\newcommand{\as}[0]{AS\=/70\xspace}

\usepackage{hyperref}
\usepackage{color}
\usepackage{cleveref}

\usepackage{pifont}

\begin{document}

\title{Multilingual Stutter Event Detection for English, German, and Mandarin Speech}

\titlerunning{Multilingual Stutter Event Detection}

\author{Felix Haas\inst{1} \and
Sebastian P. Bayerl\inst{1}\orcidID{0000-0002-3502-9511}}

\authorrunning{F. Haas}

\institute{Rosenheim Technical University of Applied Sciences, Germany \\
\email{sebastian.bayerl@th-rosenheim.de}
}

\maketitle              

\begin{abstract}

This paper presents a multi-label stuttering detection system trained on multi-corpus, multilingual data in English, German, and Mandarin.
By leveraging annotated stuttering data from three languages and four corpora, the model captures language-independent characteristics of stuttering, enabling robust detection across linguistic contexts.
Experimental results demonstrate that multilingual training achieves performance comparable to and, in some cases, even exceeds that of previous systems.
These findings suggest that stuttering exhibits cross-linguistic consistency, which supports the development of language-agnostic detection systems.
Our work demonstrates the feasibility and advantages of using multilingual data to improve generalizability and reliability in automated stuttering detection.

\end{abstract}

\section{Introduction}

Stuttering is a complex speech condition that impacts approximately 1\,~\% of individuals \cite{yairi_EpidemiologyStuttering21st_2013}.
It involves a higher frequency and longer duration of dysfluencies compared to those without this condition.
The primary symptoms of stuttering include sound, syllable, or word repetitions, prolongations, and speech blocks \cite{lickley_DisfluencyTypicalStuttered_2017}.
These may occur alongside a wide range of linguistic, physical, behavioral, and emotional symptoms.
Stuttering displays considerable variability across different individuals and within the same person over time.
Factors such as the environment, emotional state, and the linguistic demands of speech can shape how symptoms manifest \cite{ellis_HandbookStuttering_2009}.
These symptoms can hinder clear verbal expression and communication, including interactions with voice-based systems, prompting the need to identify atypical speech patterns and implement specialized systems.

Due to its dynamic nature, reliably identifying stuttering is a challenging task. 
Training data for detection systems often do not encompass the full range of their manifestations.

Previous work on stuttering detection has primarily focused on implementing various methods to enhance stuttering detection and struggled with the limited amount of available training data. 
Early work utilized neural networks for dysfluency detection on very small datasets \cite{howell_AutomaticRecognitionRepetitions_1995}.
Nöth et al. used Hidden Markov Models to detect stuttering from a proprietary dataset containing read speech samples \cite{noeth_AutomaticStutteringRecognition_2000}. 
More current works used time delay neural networks or Long Short-Term Memory (LSTM) 
\cite{sheikh_StutterNetStutteringDetection_2021,kourkounakis_DetectingMultipleSpeech_2020} to detect different types of stuttering in the UCLASS dataset \cite{howell_UniversityCollegeLondon_2009}.
A major limitation of these studies was the small datasets used. 
The detection systems trained, therefore, did not generalize well and have not been evaluated on other datasets or tasks.

The release of the \sep dataset marks a shift toward more data-oriented and less data-scarcity-oriented methods.
Showing good generalizability from methods trained on one dataset to another dataset \cite{lea_SEP28kDatasetStuttering_2021}.
They could demonstrate that dysfluency detection systems trained on one dataset generalize to another, provided sufficient training data \cite{lea_SEP28kDatasetStuttering_2021}. 
In its advent, other datasets were released, some of which were small niche application-driven datasets, such as the therapy-centered German dataset \ksof \cite{bayerl_KSoFKasselState_2022a} or the synthetically generated LibriStutter \cite{kourkounakis_LibriStutter_2021}. 
Other recent work relies on synthetically generated data \cite{zhou_YOLOStutterEndtoendRegionWise_2024}.

In related studies, it has been shown that wav2vec 2.0 (W2V2) feature extractors can be fine-tuned for stuttering detection using English data \cite{bayerl_DetectingDysfluenciesStuttering_2022a}.
The learned features were transferable from English to German for all dysfluency types except word repetitions.
Bayerl et al. \cite{bayerl_StutterSeldomComes_2023} demonstrated the utility of multilingual and cross-corpus stuttering detection systems. 
The multilingual training data improves performance on the detection task, rather than harming it, when compared to single-corpus and single-language models. 
The datasets in the study contained English and German stuttered speech. 
Both languages belong to the Germanic languages and share common properties.
Stuttering has universally shared features and traits across languages \cite{bloodstein_HandbookStuttering_2021}.
It is therefore a logical next step to incorporate all available training data to enhance robustness and promote more universal systems.
The recently released \as dataset contains Mandarin (Chinese) recordings and allows exploring the utility of multilingual training data across language families \cite{gong_AS70MandarinStuttered_2024}.

This paper utilized four datasets, \ksof, \sepE, \fb, and \as, with compatible utterance-level labels, containing five types of labeled dysfluencies. 
It examines how Mandarin stuttering data impacts the performance of multilingual stuttering detection systems that were previously trained on Germanic languages. 
For this, we use a speech-only system built upon W2V2 as described in \cite{bayerl_StutterSeldomComes_2023} and used as the baseline system in \cite{gong_AS70MandarinStuttered_2024}.
The contribution of this paper is not a new state-of-the-art dysfluency detection method, but rather a structured evaluation of the utility of multilingual training data for creating a robust multilingual detection system that generalizes across language families. 
Our contributions are 1) reproducible multi-label stuttering detection and classification using an end-to-end (E2E) system trained and evaluated on English, German, and Mandarin data; 2) experimental evidence for the generalizability and feasibility of multi-language systems across language families and datasets; 3) improving results for the \ksof dataset substantially for four out of five dysfluency types, compared to previous works.

\section{Data}\label{sec:data}

This section provides a brief description of the four datasets used and re-combined in the experiments outlined later. 
All four datasets have either no dysfluencies labeled or are labeled to belong to at least one of five types of dysfluencies, namely block (Bl), interjection (Int), prolongation (Pro), sound repetition (Snd), and word repetition (Wd). 
The \ksof dataset consists of 5597 3-second-long audio clips extracted from German stuttering therapy recordings. 
It has an additional, stuttering therapy-related label indicating the use of speech techniques learned during stuttering therapy, which we excluded for this study \cite{bayerl_KSoFKasselState_2022a}.

The Stuttering Events in Podcasts (\sep) dataset consists of $\sim$28k 3-second-long English audio clips extracted from podcasts \cite{lea_SEP28kDatasetStuttering_2021}. 

The authors of \sep also released a relabeling of the adults-who-stutter portion of the \fb \cite{bernsteinratner_FluencyBankNew_2018} dataset with compatible labels.
It contains about 4144 clips of English stuttered speech. 
Both datasets were published without suggested evaluation splits.
We therefore utilize the \sepE partitioning outlined in \cite{bayerl_InfluenceDatasetPartitioning_2022a}.

The AS-70 dataset is the largest resource containing labeled stuttered speech available for research purposes \cite{gong_AS70MandarinStuttered_2024}. 
It consists of spontaneous speech recordings derived from conversations via video conferencing software, as well as non-spontaneous read speech from voice commands. 
The dataset, unlike the others described here, is not labeled on a per-clip basis but rather uses annotations in the transcript on a character level.
For the stuttering event detection task, the authors provided a version compatible with the clip-level labels of the aforementioned datasets. 
Also, unlike the other datasets, the clips vary in length, with an average length of 4.12 seconds in the training set. 
The dataset serves as a benchmark for ASR and SED tasks on stuttered Mandarin speech \cite{gong_AS70MandarinStuttered_2024}. We use the improved partitioning from a follow-up study \cite{xue_Findings2024Mandarin_2024}.
A detailed description of the datasets and the distribution of the labels can be found in \cite{bayerl_KSoFKasselState_2022a,bayerl_InfluenceDatasetPartitioning_2022a,gong_AS70MandarinStuttered_2024,lea_SEP28kDatasetStuttering_2021}.

\section{Method}\label{sec:method}

\subsection{wav2vec~2.0}
\vspace{-2mm}
Wav2vec~2.0 \cite{baevski_Wav2vecFrameworkSelfSupervised_2020} is a family of models that utilize a convolutional feature encoder $\mathcal{G}: \boldsymbol{X}_{1:T} \mapsto \boldsymbol{Z}_{1:  L}$ composed of multiple identical layers with temporal convolution, layer normalization, and GELU activation.
This feature encoder transforms a raw audio input $\boldsymbol{X}_{1:T} = \lbrace x_1, \ldots , x_T \rbrace$ of duration $T$ into a sequence of hidden representations $\boldsymbol{Z}_{1:L} = \lbrace \boldsymbol{z}_1, \ldots , \boldsymbol{z}_L \rbrace$ of length $L$, with $L\approx\frac{T}{20ms}$.
These hidden vectors $\boldsymbol{Z}$ are then input to a transformer network, denoted as $\mathcal{T}: \boldsymbol{Z}_{1:L} \mapsto \boldsymbol{C}_{1:L}$, which produces contextualized outputs $\boldsymbol{c}_1, \ldots , \boldsymbol{c}_L$.
During pre-training, a quantization module discretizes $\boldsymbol{Z}_{1:L}$, and the model is trained using a contrastive objective to identify the correct quantized vector among distractors.
Representations extracted from various transformer layers of wav2vec~2.0 are effective acoustic features for detecting dysfluencies \cite{bayerl_StutterSeldomComes_2023,bayerl_DetectingDysfluenciesStuttering_2022a} and for related applications like mispronunciation detection \cite{baevski_Wav2vecFrameworkSelfSupervised_2020,xu_ExploreWav2vecMispronunciation_2021}.
Pre-trained W2V2 models are available in two sizes (94M and 315M parameters). 

We use the large W2V2 models in our experiments, largely following \cite{bayerl_StutterSeldomComes_2023} to maintain consistency and comparability.
The weights for the models used in the experiments were pre-trained on roughly 4.5M hours of audio in 53 languages \footnote{\url{https://huggingface.co/docs/transformers/model_doc/xlsr_wav2vec2}}, or pre-trained on 10k hours of the WenetSpeech L subset \footnote{\url{https://huggingface.co/TencentGameMate/chinese-wav2vec2-large}}.

Unlike the approach described in \cite{bayerl_StutterSeldomComes_2023}, which uses an attention-based pooling mechanism, we use the standard W2V2 sequence classification head with mean pooling over time.
The model was altered from the standard implementation outlined in \cite{wolf_TransformersStateoftheArtNatural_2020} in only two key ways:
Firstly, the architecture supports multi-task learning through the addition of an auxiliary output branch, following the design in \cite{bayerl_DetectingDysfluenciesStuttering_2022a}.

Secondly, instead of using a softmax classifier on the main output branch, we apply an element-wise sigmoid function ($\sigma$), allowing the model to estimate the likelihood of each class independently for a given audio segment.

\vspace{-2mm}
\subsection{Loss}\label{ss:mtl}
\vspace{-2mm}
Multi-task learning (MTL) is a common regularization technique that helps prevent overfitting and improves generalizability, especially when the auxiliary task supports the main task. 

MTL is widely applicable across speech tasks \cite{pironkov_SpeakerawareLongShortterm_2016,ravanelli_MultiTaskSelfSupervisedLearning_2020,cai_SpeechEmotionRecognition_2021}.

Previous studies on stuttering detection have conclusively shown that MTL improves the performance of such systems. 
Auxiliary tasks explored in previous studies on stuttering detection have ranged from using an artificial `any' label, which indicates the presence of any dysfluency in a speech clip, to gender classification and language detection \cite{lea_SEP28kDatasetStuttering_2021,bayerl_DetectingDysfluenciesStuttering_2022a,sheikh_RobustStutteringDetection_2022,bayerl_StutterSeldomComes_2023}. 
In our experiments, we exclusively use language detection as an auxiliary task.

In all experiments outlined in this paper, we use a weighted combination of Binary Cross Entropy (BCE) and Focal Loss (FL) \cite{lin_FocalLossDense_2020}. 
FL is an extension of the BCE loss using the $\alpha$ and $\gamma$ parameters to emphasize the importance of minority classes and handle class imbalance. 

\begin{equation}\label{eq:focal_loss}
\vspace{-2mm}
    \mathbf{L_{FL}}(p_t) = -\alpha (1 - p_t)^\gamma \log{(p_t)}.
\end{equation}

FL can be used as a drop-in replacement for the BCE loss in multi-label classification tasks by calculating the loss individually for each class directly on the output of each neuron. 
In our experiments, the overall FL is calculated by computing the mean loss value for each class as shown in \cref{eq:focal_loss_multi}.

\begin{equation}\label{eq:focal_loss_multi}
\vspace{-2mm}
    \mathbf{L_{{FLmulti}}} = \frac{1}{n}\sum^n_1{\mathbf{FL_n}(p_t)}
\vspace{-2mm}
\end{equation}

The final loss is computed by combining both the BCE loss calculated for the auxiliary task ($L_{\text{aux}}$) and the FL calculated for the main task ($L_{\text{main}}$) 
as a weighted sum, with $w_{main}$ as a hyperparameter weighing the contribution of the tasks, as shown in \cref{eq:mtlloss}.

\begin{equation}\label{eq:mtlloss}
\vspace{-2mm}
    \mathbf{L}_{\text{MTL}} = w_{\text{main}} L_{\text{main}} + (1-w_{\text{main}}) L_{\text{aux}}
\end{equation}

\section{Experiments}\label{sec:experiments}

All experiments described in this section were performed using the general setup described in \cref{sec:method}.
The model training was performed on different combinations of training data. 
Preliminary experiments, which were too comprehensive to be reported here in full, were used to determine the hyperparameters and the composition of the training data, as well as the weights of the pre-trained W2V2 models that were chosen among English, Chinese, and XLSR weights.

The best overall results were achieved using the XLSR-53 weights, which utilize pre-training data from 53 languages, including German, Chinese, and English.

The loss was scaled and then weighted. Preliminary experiments determined $w_{main}=0.9$. 
The parameters for the FL were set to $\alpha = 0.7$ and $\gamma = 3$ 
All models were trained using the AdamW optimizer, a batch size of 8, and a learning rate of $lr=\num{3e-5}$. 
Training was performed for up to 20 epochs with early stopping implemented with a patience of 5.

\subsubsection{Inference on \as:}
As the \as dataset, unlike the other datasets, consists of variable-length audio sequences, inference on the longer clips is a composition of multiple inference steps that are aggregated to the clip level.
This means splitting sequences longer than \SI{3}{\second} into multiple 3-second-long clips with a 1.5-second overlap.

Inference for the sequences is run individually, and detection results for each clip are aggregated and treated as one sample.

\begin{figure}[!ht]
    \centering
    \includegraphics[width=0.8\linewidth]{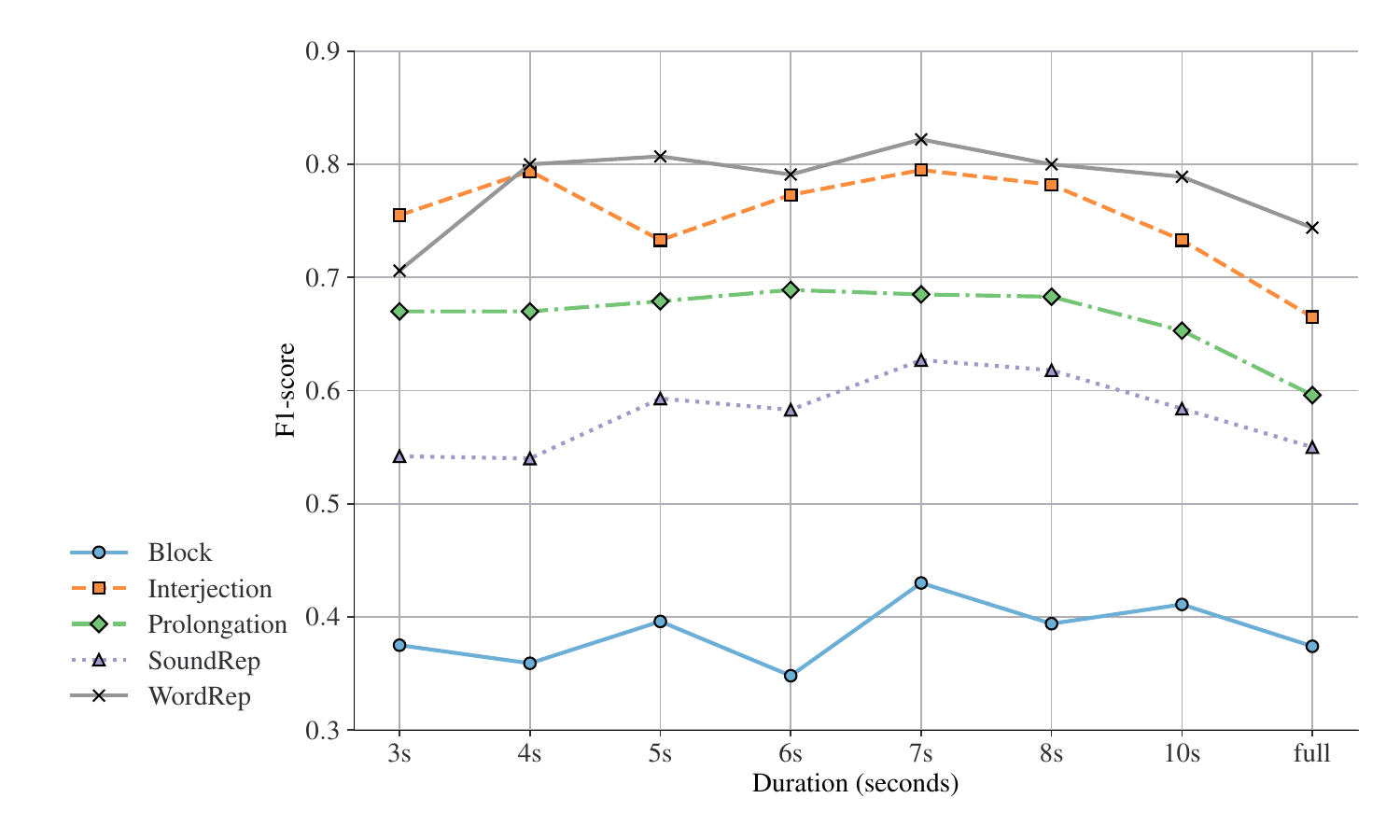}
    \caption{$F_1$ scores relative to the maximum clip lengths included in the training data.}
    \label{fig:length-perf-plot}
\end{figure}
\subsection{Experimental setup}

For the experiments, models were trained solely on the training portions of the four datasets described in \cref{sec:data}. 

\subsubsection{Baseline}
Baseline results are reported from the papers \cite{bayerl_StutterSeldomComes_2023} and \cite{gong_AS70MandarinStuttered_2024}.
For the result \#4 in \Cref{tab:xling_results_data}, we trained a model based on the combination of the training sets of \ksof, \fb, and \sepE, denoted as \textit{"Multi-Lingual"} in \cite{bayerl_StutterSeldomComes_2023} and evaluated it on the \as test set. 
We will refer to this system as EN-DE going forward. 

\subsubsection{\as}

This experiment recreates the W2V2 baseline reported in \cite{gong_AS70MandarinStuttered_2024}. 
The model was trained on all clips in the \as the training set. 

In addition to the \as test set, it was also evaluated on the test sets of \ksof, \fb, and \sepE to assess the performance on German and English data of models trained solely on Mandarin data.

\subsubsection{TRILANG}

For the TRILANG training, the training data of all four datasets, \ksof, \sepE, \fb, and \as, were combined and shuffled. 
The longest sequences after the pre-processing scripts provided by the authors were 14.375 seconds long.

\subsubsection{TRILANG-LL}
We hypothesized that mixing \SI{3}{\second}-long clips from \ksof, \sepE, and \fb with substantially longer sequences from \as could negatively affect overall model performance. 
To investigate this, we created variants of the training set by applying different maximum length thresholds to \as (\SI{3}{\second} to no threshold).
The composition of the test set remained unchanged and consistent across all experiments.
The results of these preliminary experiments are visualized in \Cref{fig:length-perf-plot}.

\begin{figure}[htb]
    \centering
    \includegraphics[width=0.7\linewidth]{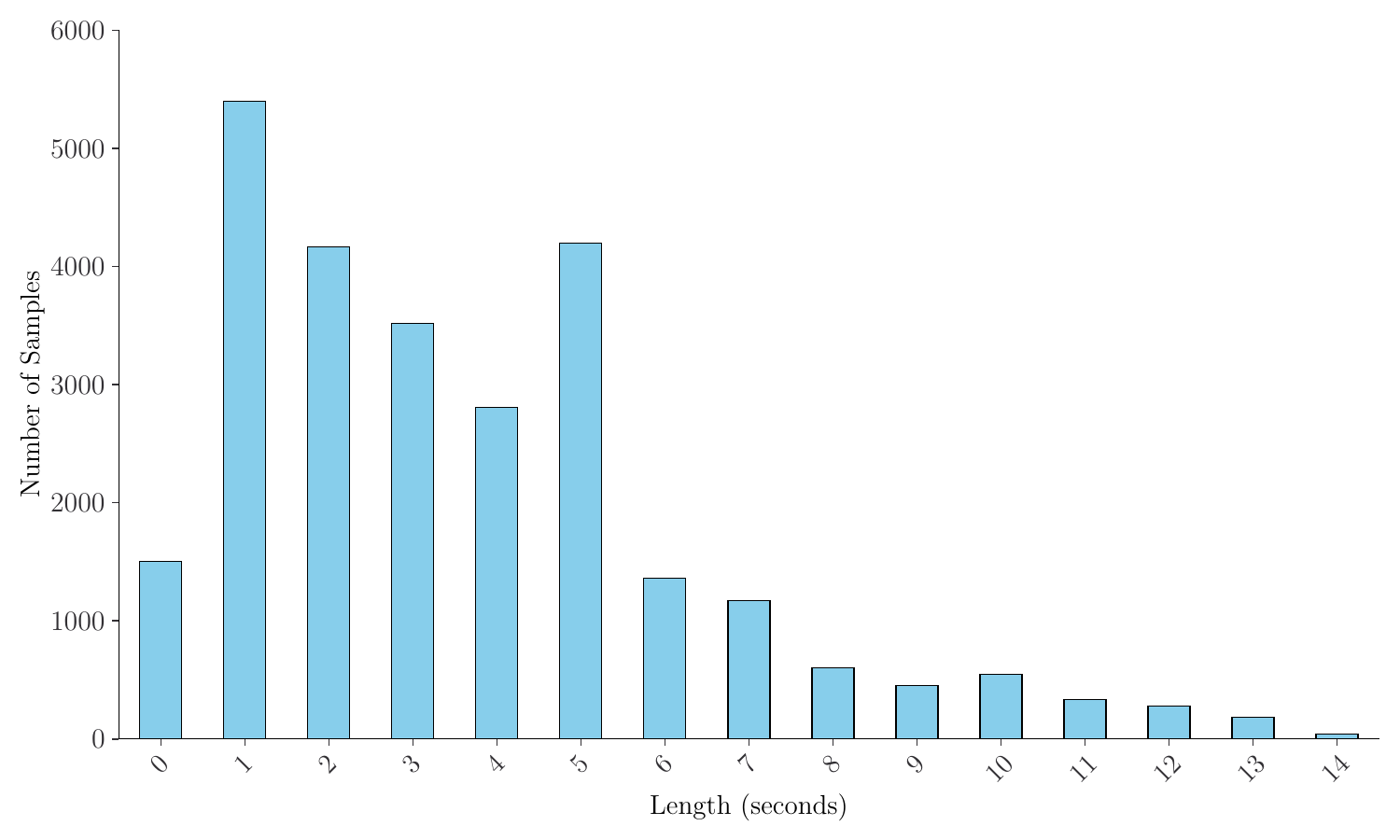}
    \caption{Distribution of clip lengths in the \as dataset.}
    \label{fig:length-dist}
    \vspace{-4mm}
    
\end{figure}

TRILANG-LL refers to the length-limited subset of the training data.
Preliminary experiments indicated that the optimal threshold for sequence length was at \SI{7}{\second}. 
Applying this threshold retained approximately 86~\% of the clips and about 69~\% of the total duration of the original training data.
The length distribution of the clips is shown in \cref{fig:length-dist}.

\subsection{Results}

This section briefly describes the experimental results summarized in \Cref{tab:xling_results_data}.

Using a system similar to that of \cite{bayerl_StutterSeldomComes_2023} proves effective. 
The results presented in \cref{tab:xling_results_data} \#9 show a substantial improvement compared to those in \#4, which were obtained using inference with the EN-DE model.
The baseline performance reported in \cite{gong_AS70MandarinStuttered_2024} could not be reproduced using our MTL system, with results falling short by a noticeable margin (comp. \cref{tab:xling_results_data} \#~5 and \#~9).

Applying the model trained on only \as to the other datasets (\cref{tab:xling_results_data} \#~6 to \#~8) leads to results that are better than chance predictions. 
However, performance across the board falls short of that achieved using the EN-DE trained model.
Despite stuttering exhibiting similar patterns across languages, the model struggles to capture the linguistic intricacies of a different language. This leads to comparably poor results as using the EN-DE model on the Mandarin dataset (comp. \#~4).
Furthermore, performance on the AS-70 dataset (comp. \#~9) is slightly disappointing, as it falls short of the baseline reported in \cite{gong_AS70MandarinStuttered_2024}.

The TRILANG training yields results for \ksof, \sep, and \fb that exhibit a positive trend, nearly matching the performance of the EN-DE model (comp. \cref{tab:xling_results_data} \#~1--3 to \#~10--12).
Notably, \textbf{Wd} improves for both \fb and \ksof, even surpassing the baseline, and almost matches the performance for \sep.
Regarding the recognition of \textbf{Snd}, the performance improves over the baseline for \ksof, and slightly worsens for \fb and \sep.
However, all remain within a reasonable distance to the baseline performance.
For \textbf{Pro}, the TRILANG model matches the performance on \ksof compared to the EN-DE system, while performance on \sep and \fb falls slightly behind.
The detection results for \textbf{Int}s show a clear improvement over the baseline for the English datasets.
\textbf{Bl} detection remains weaker overall, yet all three datasets (\fb, \ksof, \sep) show improved performance compared to models trained solely on Chinese data.

The TRILANG model for Mandarin demonstrates that adding English and German training data to the training set causes little to no change to the performance on \as, compared to the model trained on \as only (see \#~9,~13).
The most notable drop in performance is observed for the detection of Mandarin \textbf{Pro}, where the multilingual training appears to have a detrimental effect.

\begin{figure}
    \centering
    \includegraphics[width=0.6\linewidth]{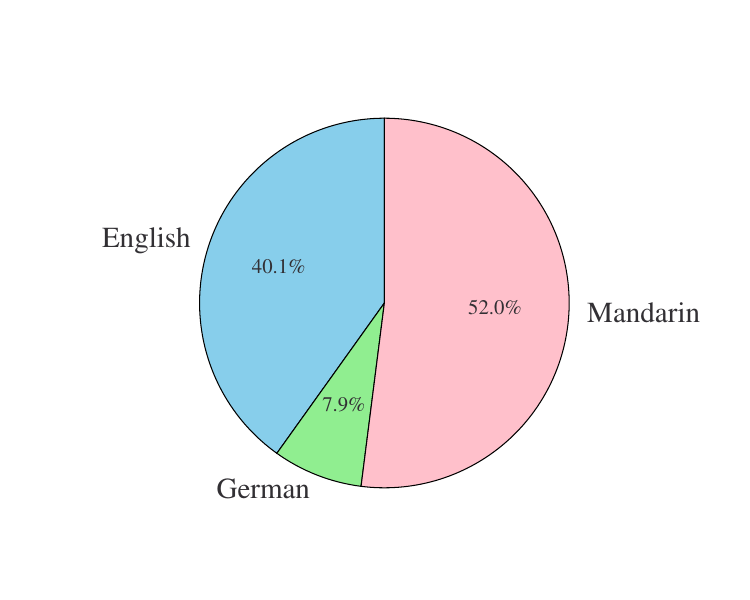}
    \caption{Distribution in \# of distinct samples in the TRILANG-LL training set.}
    \vspace{-4mm}
    \label{fig:trilang-lang-dist}
\end{figure}

The TRILANG-LL trained model yields improvements across all evaluated datasets.
Most notably, \textbf{Wd} detection shows consistent gains, including for the Mandarin data, which substantially improves over the baseline and all other models (\cref{tab:xling_results_data} \#~14--17).  
The overall performance boost compared to both the \as-only training setup (comp. \#9) and the standard TRILANG configuration stands out.
The TRILANG-LL model reaches or improves performance for \textbf{Wd}, \textbf{Snd}, and \textbf{Int} over the baseline for all three non-Mandarin datasets. 
The improvements for the German \ksof dataset are most striking, as the TRILANG-LL trained model outperforms the baseline for all but \textbf{Bl}, where it still reaches comparable performance.
This is particularly interesting, as German has the smallest share of the overall training data (see \cref{fig:trilang-lang-dist}).

\begin{table}[!t]
	\centering
	
	\caption{
		Cross-Dataset and multi-lingual dysfluency detection results ($F_1$-score) using E2E multi-label systems.
		(\textbf{Bl} = Block, \textbf{Int} = Interjection, \textbf{Pro} = Prolongation, \textbf{Snd} = Sound repetition, \textbf{Wd} = Word repetition).
		Table section headers indicate the training data used, as described in \cref{sec:experiments}, followed by the W2V2 base-weights used.
		\textbf{Bold} values indicate the highest $F_1$ score for a dysfluency type in one of the datasets.
	}

	\begin{tabular}{c|c|c|c|c|c|c}

		\toprule
		\# & \textbf{Test-Set}                                                                                                                             & \textbf{Bl}   & \textbf{Int}  & \textbf{Pro}  & \textbf{Snd}  & \textbf{Wd}   \\
		\midrule
		\multicolumn{7}{c}{\textbf{Baseline}}                                                                                                                                                                                              \\
		\midrule
		1  & \sepE \cite{bayerl_StutterSeldomComes_2023}                                                                                                   & 0.32          & \textbf{0.77} & \textbf{0.53} & \textbf{0.53} & 0.64          \\
		2  & \fb \cite{bayerl_StutterSeldomComes_2023}                                                                                                     & \textbf{0.36} & 0.79          & \textbf{0.62} & 0.64          & 0.52          \\
		3  & \ksof \cite{bayerl_StutterSeldomComes_2023}                                                                                                   & \textbf{0.64} & 0.85          & 0.60          & 0.48          & 0.14          \\
		4  & \as \tablefootnote{inference with a model trained using the ``Multi-Lingual'' training data from \cite{bayerl_StutterSeldomComes_2023}}   & 0.28          & 0.61          & 0.60          & 0.47          & 0.56          \\
		5  & \as \cite{gong_AS70MandarinStuttered_2024}                                                                                                    & \textbf{0.43} & \textbf{0.84} & \textbf{0.70} & \textbf{0.66} & 0.78          \\
		\midrule
		\multicolumn{7}{c}{\textbf{\as (XLSR-53)}}                                                                                                                                                                                         \\
		\midrule
		6  & \sepE                                                                                                                                         & 0.22          & 0.46          & 0.34          & 0.27          & 0.28          \\
		7  & \fb                                                                                                                                           & 0.15          & 0.56          & 0.39          & 0.40          & 0.36          \\
		8  & \ksof                                                                                                                                         & 0.23          & 0.52          & 0.45          & 0.36          & 0.15          \\
		9  & \as                                                                                                                                           & 0.37          & 0.68          & 0.64          & 0.56          & 0.73          \\
		\midrule
		\multicolumn{7}{c}{\textbf{TRILANG (XLSR-53)}}                                                                                                                                                                                     \\
		\midrule
		10 & \sepE                                                                                                                                         & 0.26          & \textbf{0.77} & 0.52          & 0.47          & 0.61          \\
		11 & \fb                                                                                                                                           & 0.26          & 0.82          & 0.60          & 0.62          & 0.58          \\
		12 & \ksof                                                                                                                                         & 0.55          & 0.81          & 0.60          & 0.51          & 0.32          \\
		13 & \as                                                                                                                                           & 0.37          & 0.67          & 0.60          & 0.55          & 0.74          \\
		\midrule
		\multicolumn{7}{c}{\textbf{TRILANG-LL (XLSR-53)}}                                                                                                                                                                                  \\
		\midrule
		14 & \sepE                                                                                                                                         & \textbf{0.33} & \textbf{0.77} & 0.51          & \textbf{0.53} & \textbf{0.71} \\
		15 & \fb                                                                                                                                           & 0.32          & \textbf{0.84} & 0.61          & \textbf{0.66} & \textbf{0.59} \\
		16 & \ksof                                                                                                                                         & 0.63          & \textbf{0.88} & \textbf{0.66} & \textbf{0.55} & \textbf{0.35} \\
		17 & \as                                                                                                                                           & \textbf{0.43} & 0.80          & 0.69          & 0.63          & \textbf{0.82} \\
		\bottomrule
	\end{tabular}

	\label{tab:xling_results_data}

\end{table}

\subsection{Discussion}

Different stuttering symptoms rely on distinct temporal contexts for their detection.
\textbf{Snd} and \textbf{Pro} typically require only short contextual windows, as they occur within a single word or syllable. 
\textbf{Int} span slightly longer durations—often encompassing partial or complete words.
\textbf{Bl} demand even longer temporal context, as they must be distinguished from prosodic pauses, which are common in fluent speech. 
\textbf{Wd} are challenging, as they require a long contextual window. 
Since \textbf{Wd} can involve a wide range of lexical items, many of which are low-frequency, it is unlikely that all test-set repetitions are represented in the training data.

This contrasts with sound repetitions and prolongations, which involve a much smaller set of units.
For instance, English has 44 phonemes, while Mandarin has 26 phonemes, with added complexity due to tonal distinctions \cite{duyen_ExploringPhoneticDifferences_2024}.
Nevertheless, it is reasonable to assume that all relevant phonetic units were encountered during training for these symptoms.
Thus, word repetitions can be considered a type of meta-pattern: a long-context repetition phenomenon. 
We therefore hypothesize that the improvements in \textbf{Wd} detection stem from the model effectively learning this meta-pattern only when exposed to more diverse training data, even if such data originates from a different language family.

Another interesting aspect is the strong performance on the \ksof data. 
While only 7.9~\% of the training data was German, the TRILANG-LL model trained on English, Mandarin, and German, surpasses the previous state of the art.
It seems very little data in a specific language is needed to generalize towards it, which is not entirely surprising, as stuttering shares common traits such as frequent repetitions of plosives or prolongations of nasals and vowels. 

The decision to apply a length limitation was empirical and motivated by practical considerations.
While longer context windows could, in theory, benefit model performance, the training data contains relatively few long sequences—only around 14\% of the clips in \as exceed the \SI{7}{\second} threshold, and these occur in only one language. 
Moreover, longer clips tend to exhibit a greater number of dysfluencies, often of multiple types, increasing the complexity of the classification task.
Including such longer sequences introduces potential sources of noise, such as padding artifacts or information loss due to truncation. 
These factors can negatively impact training efficiency and generalization.

Upon further analysis, it was observed that filtering out unlabeled clips reveals an average of 1.66 dysfluencies per clip with a \SI{3}{\second} threshold. 
Only 169 clips longer than \SI{7}{\second} contained no dysfluencies at all. 
Clips exceeding \SI{7}{\second} contain, on average, more than two dysfluencies, which increases with higher thresholds, as expected.
However, the disfluency labels do not indicate whether the same type appears multiple times within a clip. 
This makes it challenging to generalize effectively, and it further justifies excluding long clips to maintain clarity and consistency in the training data, providing a potential explanation for why performance improves when these longer clips are excluded.

\vspace{-4mm}

\subsubsection{Limitations}

A key limitation of our approach is the coarse prediction granularity for longer audio sequences, which are split into overlapping segments. 
This inevitably leads to the model seeing the same stuttering event in different contexts and having more opportunities for detection in the case of the longer clips in the \as data.
Examining the consistent $F_1$ scores across conditions, it becomes clear that the observed performance improvements—or the maintenance of performance levels—are not solely due to an increased recall (see exp. \# 9, 13, 17). 
Precision does not degrade; if it did, we would expect a corresponding decline in $F_1$ scores. 
This suggests that the balance between precision and recall is being preserved, indicating genuine improvement rather than a trade-off between the two metrics.
As a consequence, the prediction is reduced to the clip level, obscuring the location of stuttering events within longer utterances. 
This compromise, however, is in line with the constraints defined by the \as clip-level labels and represents a practical trade-off.

Another limitation is the reliance on an older model architecture while more advanced models, such as WavLM \cite{chen_WavLMLargeScaleSelfSupervised_2022}, Whisper \cite{radford_RobustSpeechRecognition_2022}, or BEST-RQ \cite{chiu_SelfsupervisedLearningRandomprojection_2022} have been proposed since.
This decision is intentional.
Maintaining consistency with prior methods helps isolate the effect of linguistic variation.
This allows us to attribute performance differences more confidently to the addition of data from a new language family rather than to architectural improvements. 
It is reasonable to assume that scaling laws in representation learning apply broadly, making our findings relevant to newer architectures.

\FloatBarrier
\vspace{-2mm}
\section{Conclusion}
\vspace{-2mm}

This paper examined the effects of combining labeled stuttering speech data from English, German, and Mandarin to train a multi-label, cross-lingual disfluency detection system using W2V2. 
The results demonstrate that leveraging multilingual training data does not degrade performance and even has the potential to improve detection of complex repetition patterns, likely due to greater linguistic diversity in the training corpus. 
The amount of training data remains a critical factor in building more effective stuttering detection systems.
In the era of large language models, multilingual systems have become the norm. 
Speech recognition systems such as Whisper \cite{radford_RobustSpeechRecognition_2022} are capable of handling multiple languages, and future stuttering detection systems should follow.

While the usefulness of synthetic data is often overstated, its potential should be further explored \cite{kourkounakis_LibriStutter_2021,zhou_YOLOStutterEndtoendRegionWise_2024}.
It might be particularly useful for pre-training on low-resource languages.
As demonstrated in this paper, only a relatively small amount of data was needed to adapt the model for German stuttering effectively.

Future work will focus on incorporating synthetic data and expanding to additional languages using new resources such as the Boli dataset \cite{batra_BoliDatasetUnderstanding_2025}, which includes stuttering speech in several Indian languages.

\bibliographystyle{splncs04}
\footnotesize{
\bibliography{zotero_no_url}
}

\end{document}